\documentclass[12pt]{iopart}
\usepackage{booktabs}
\usepackage{graphicx}% Include figure files
\usepackage{dcolumn}% Align table columns on decimal point
\usepackage{bm}% bold math
\usepackage{subfigure}
\usepackage{indentfirst}
\begin{document}

\title[A quantitative method]{A quantitative method for determining the robustness of complex networks}
%复杂网络的鲁棒性测度
\author{Jun Qin$^{1,2}$, Hongrun Wu$^1$, Xiaonian Tong$^1$,
Bojin Zheng*$^{1,2,3}$}

\address{$^1$ College of Computer Science,
South-Central University for Nationalities, Wuhan 430074, China}
\address{$^2$ State Key Laboratory of Networking and Switching Technology, Beijing University of Posts and Telecommunications, Beijing 100876, China}
\address{$^3$ School of Software, Tsinghua University, Beijing 100084, China}
\ead{zhengbojin@gmail.com}

\begin{abstract}
 Most current studies estimate the invulnerability of complex networks using a qualitative method that analyzes the decay rate of network performance. This method results in confusion over the invulnerability of various types of complex networks. By normalizing network performance and defining a baseline, this paper defines the invulnerability index as the integral of the normalized network performance curve minus the baseline. This quantitative method seeks to measure network invulnerability under both edge and node attacks and provides a definition on the distinguishment of the robustness and fragility of networks.  To demonstrate the proposed method, three small-world networks were selected as test beds. The simulation results indicate that the proposed invulnerability index can effectively and accurately quantify network resilience and can deal with both the node and edge attacks. The index can provide a valuable reference for determining network invulnerability in future research.
\end{abstract}

\maketitle

\section{Introduction}
Recently, the work of the pioneering scientists Albert et al. on network resilience has attracted much attention \cite{315}. Based on the work of Albert et al. , Holme et al. subsequently extended the concepts of (node) degree and betweenness to edge degree and edge betweenness, and explored the invulnerability of complex networks under edge attacks \cite{146}. Furthermore, Motter \cite{316}, Holme \cite{317} and other researchers have investigated the dynamic effects of complex networks under node and edge attacks. Song et al. have also discussed the fragility of fractal networks \cite{145}. The real-world networks, such as metabolic networks \cite{310}, scientific cooperation networks \cite{80}, social networks \cite{8,19,20}, and email networks \cite{34} etc. have also been investigated \cite{308}.  However, in most of these studies, network invulnerability was analyzed by qualitatively assessing the decay rate of network performance \cite{315,317,152}.

Recent studies have shared new perspectives in understanding network resilience with the quantitative way. Schneider et al. \cite{318} defined a quantitative index -- $R$ index, and used this index to validate a conclusion that the scale-free networks with an onion structure are the most robust scale-free networks against malicious attacks, although still fragile \cite{178}. Later, Schneider et al. \cite{238} extended the $R$ index to deal with the edge attacks. Notice that the node attacks can be transferred into the edge attacks, the index values should be same or close when calculating the node attacks and the corresponding edge attacks. However, the $R$ index does not pay enough attention on this constraint.

The current paper tries to provide a quantitative method to be able to address both node and edge attacks, and have close values when dealing with certain node attacks and the corresponding edge attacks. Moreover, this method prefers to provide a definition on a clear distinguishment to the robustness and fragility of networks. To meet these aims, we first normalize the network performance using the normalized size of the removed edges and define a baseline as a benchmark for distinguishing robustness and fragility of the network under an attack. Second, to quantify the network invulnerability under an attack strategy, i.e. , a sequence of attacks, we define the invulnerability index as the integral of the normalized network performance curve minus the baseline. This method can assure the approximated index values when dealing the equivalent node and edge attacks.

To demonstrate the proposed invulnerability index, we explore the robustness of small-world networks. Two small-world networks from the real world and one WS model network \cite{16} are tested under edge and node attacks. The simulation results demonstrate that the proposed method is effective at quantifying network resilience under node and edge attacks and at distinguishing network robustness and fragility.

\section{The quantitative method}

The proposed method aims to 1) address both node and edge attacks and  2) provide a definition of the clear and quantitative demarcation between robustness and fragility. To achieve the first aim, the proposed method suggests normalizing the network performance using the normalized size of the removed edges. Because node attacks can be regarded as attacks on bundles of edges, this method is capable of simultaneously addressing node and edge attacks with the constraint that the network under certain node attacks and corresponding edge attacks will have approximated values. To provide a clear and quantitative distinction between robustness and fragility under an attack, this method defines a continuous curve --- the baseline, and then defines the invulnerability index as the integral of the normalized network performance curve minus the baseline, then establishes a benchmark such that if the invulnerability index of a network is greater than 0, then the network is robust; otherwise, the network is fragile.

\subsection{Normalized network performance}

Previous studies have proposed many methods for measuring network performance \cite{34,146}. In this paper, we use the size of the giant component of a given graph to determine network performance after a set of edges has been removed. When dealing with node attacks, the removal of nodes is regarded as the removal of bundles of edges. For example, the removal of node $A$ can actually be regarded as the removal of all edges of node $A$.

Assume that an initial network has $N$ nodes and $E$ edges. After the removal of a set of edges $T$, the size of the giant component is $\tilde{s}(T)$. The normalized network performance with respect to the normalized size of the removed edges $r$ can be expressed as equation (\ref{eq1}). In this equation, $s(r)\in[0,1]$, and $r \in[0,1] $.

\begin{equation}\label{eq1}
    s(r)= \frac{\tilde{s}(T)}{N}
\end{equation}

where
\begin{equation}\label{eq2}
    r= \frac{|T|}{E}
\end{equation}

Because the number of edges is an integer (i.e., is discrete), any function of the normalized network performance is also discrete. To define a baseline and compute the invulnerability index, the function of the normalized network performance should be continuous. We use the most common method for creating this function: i.e., linking adjacent discrete points one by one to form a network performance curve. This method is helpful of the approximation of the index values between the node attacks and the corresponding edge attacks.

However, one problem associated with node attacks should be mentioned. The removal of a node is regarded as the removal of a bundle of edges. Because the sequence in which the edges are removed is uncertain, the normalized network performance varies.

To simplify this problem, this paper suggests a method of interpolation.

%Noticed that the network performance values are certain before and after the removal of a node,this paper assigns values to the middle network performance values.

Assume that the network performances before and after node removal are $s_i$ and $s_j$, respectively; Here, $i$ and $j$ are the number of the edges of the giant component. Thus, the network performance after the removal of $m$ edges is defined by equation (\ref{eq.per}).

\begin{equation}\label{eq.per}
    s_m=s_i + \frac{m}{j - i} (s_j - s_i) \ \ \ (0 \leq m \leq j-i)
\end{equation}

Using this method of interpolation, the node attacks are generalized to edge attacks, and the network performance under node attacks approximates the network performance under edge attacks.

%%提出常用的测度标准
%%前期研究中，有很多指标来评估网络鲁棒性。
%%在本文中，我们将给出几个常用的指标来评估网络鲁棒性

\subsection{Definition of baseline}

When a network is randomly or maliciously attacked, a coordinate is often established to characterize the relationship between the attack and the network performance (see Fig. 1). In Fig. 1, the horizontal axis represents the proportion of edges removed under each attack, and the vertical axis represents the normalized network performance. Each attack and the corresponding network performance is plotted as a point. After a sequence of attacks, the relationship between the attacks and the network performance is plotted as a discrete curve.

\begin{figure}[htbp]\label{fig1}
\centering\includegraphics[width=8cm,height=6cm]{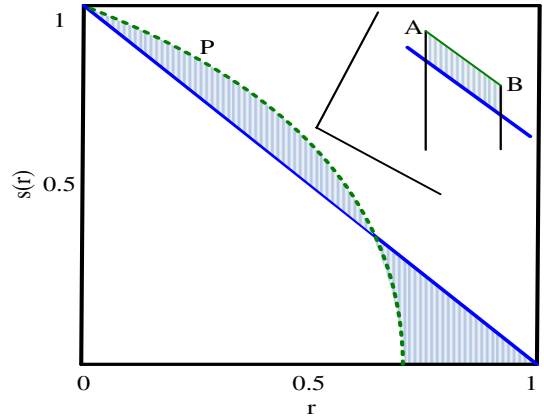}% Here is how to import EPS art
\caption{Measurement of network invulnerability. The vertical axis $s(r)$ is the normalized network performance, and the horizontal axis $r$ represents the fraction of removed edges. The continuous blue diagonal line is the baseline; the green dashed curve is the discrete network performance curve, and P is a point on the discrete curve. When calculating the invulnerability index, the discrete network performance curve should be continuous. The discrete adjacent points A and B are continued as curve AB. The shaded regions are the difference between the baseline and the network performance curve. The area above the baseline minus the area below the baseline is the value of the invulnerability index.}
\end{figure}

Most current methods define network invulnerability by considering the decay rate of network performance. In this paper, we set a baseline so as to more easily recognize the robustness or the fragility in attack graphs and to mark the critical point at which the fragility of a graph emerges. The baseline is shown as the blue diagonal curve in Fig. 1 and evenly divides the area of the coordinates into two parts.

This baseline is an equilibrium between the removed edges and the lost nodes. When a fraction, for example, 5\%, of edges are removed, there are three results for the lose of the nodes. (1) 5\% nodes are removed; (2) less than 5\% nodes are removed; (3) more than 5\% nodes are removed. For result (1), the relationship of the fraction of the removed edges and the responding network performance is linear, we define it as the baseline. As to result (2), the curve is plotted above the baseline, defined as robust, and for result (3), the curve is below the baseline, defined as fragile.

Mathematically, the baseline can be formalized as equation (\ref{eq.baseline}).

\begin{equation}\label{eq.baseline}
    f(r)=1-r
\end{equation}

%************************************************

The baseline is used as a strict demarcation to distinguish the network robustness from the fragility in the attack graph. When a point on the network performance curve is above the baseline, the network is robust to a given attack; otherwise, it is fragile. For example, point P in Fig. 1 is above the baseline under a given attack; therefore, the network is robust to this attack. Correspondingly, when a sequence of points are above the baseline, the network is robust to a given series of attacks. Otherwise, the network is fragile.

However, when some points are above the baseline and some under the baseline, a demarcation is needed. Moreover, a more accurate demarcation of the invulnerability of the entire network is also required. Thus, this paper uses the integral of the normalized network performance curve  minus the baseline to quantify the network invulnerability.

\subsection{The invulnerability index }

Based on the normalized network performance curve and the baseline, this paper defines the invulnerability index $I$ as the integral of the normalized network performance curve minus the baseline. When a certain fraction of edges (denoted as $\alpha$) are removed, the invulnerability index is denoted as $I_{\alpha}$.
Formally, $I_{\alpha}$ is defined as equation (\ref{eq.generalIndex}).
%and the regions above the baseline contributes positive area, whereas the regions under the baseline contributes negative area.

%i.e., the difference between the bounded area by the normalized network performance curve and the bounded area by the baseline.

\begin{equation}\label{eq.generalIndex}
I_{\alpha}= \int_0^\alpha (s(r)-f(r))\,dr = \int_0^\alpha (s(r)-1+r)\,dr
\end{equation}

Equation (\ref{eq.generalIndex}) can be rewritten as equation (\ref{eq.generalIndex2}).

\begin{equation}\label{eq.generalIndex2}
I_{\alpha}= \int_0^\alpha s(r)\,dr + (\frac{\alpha^2}{2}- \alpha)
\end{equation}

The key to solving equation (\ref{eq.generalIndex2}) is calculating the integral $\int_0^\alpha s(r)\,dr$ of the normalized network performance curve. In fact, here the integral $\int_0^\alpha s(r)\,dr$ is the area bounded by the normalized network performance curve and the coordinate axes.

Because $E$ is the number of edges, the discrete normalized network performance function $s(r)$ has $E + 1$ points \{ $s_0, s_1, ..., s_E$\}, including the initial graph without any attacks. Therefore, the area $A_{\alpha}$ bounded by the normalized network performance curve and the coordinate axes satisfies equation (\ref{eq.22}) when a percentage $\alpha$ of the edges are removed as follows:

\begin{equation}\label{eq.22}
A_{\alpha}=(\frac{s_0+s_1}{2}+ \frac{s_1+s_2}{2} + ... +  \frac{s_{e-1}+s_e}{2})*\frac{1}{E}
\end{equation}

where
\begin{equation}\label{eq.e}
e=\lceil\alpha*E\rceil
\end{equation}

Thus, equation (\ref{eq.generalIndex2}) can be rewritten as equation.

\begin{equation}\label{eq.generalIndex3}
I_{\alpha}= A_{\alpha} + (\frac{\alpha^2}{2}- \alpha)
\end{equation}

Because the domain of the coordinates is $[0,1]$, so $0 \leq A_{\alpha} \leq 1$. Therefore, $I_{\alpha} \in [-0.5, 0.5]$.

Thus, the invulnerability index can be quantified. In extreme cases, the network performance curve may completely overlap the baseline, $I_{\alpha}=0$. When the area above the baseline bounded by a fraction of points, $\alpha$, is smaller than the area under the baseline, $I_{\alpha} < 0$, the network is fragile; otherwise, $I_{\alpha} > 0$, and the network is robust. With the same fraction of removed edges, a larger invulnerability index indicates a more robust network.

In particular, when all of the edges are removed from a network, the invulnerability index is represented by equation (\ref{eq.index21}), and the index $I_{1}$ is often used by researchers to estimate network resilience.

\begin{equation}\label{eq.index21}
I_{1}= \int_0^1 (s(r)-1+r)\,dr
\end{equation}

When $\alpha$ = 1, we can calculate the invulnerability index using a simple method.
According to geometry, the area bounded by the baseline and the coordinate axes is always a constant $1/2$, therefore, $I_{1}$ is the area bounded by the network performance curve and the coordinate axes minus the constant $1/2$, which is written as equation (\ref{eq.boundedarea}).
\begin{equation}\label{eq.boundedarea}
I_{1}= A_{1}-\frac{1}{2}
\end{equation}

where
\begin{equation}\label{eq.a1}
A_{1}=(\frac{s_0+s_1}{2}+ \frac{s_1+s_2}{2} + ... +  \frac{s_{E-1}+s_E}{2})*\frac{1}{E}
\end{equation}

Note that $s_0 = 1$ and $s_E = 0$; thus, equation (\ref{eq.a1}) can be rewritten as equation (\ref{eq.A2}).
\begin{equation}\label{eq.A2}
A_1=(\sum\limits_{i = 0}^E {s_i } - \frac{1}{2})* \frac{1}{E}
\end{equation}

The invulnerability index $I_{1}$ only reflects the final status of the robustness or fragility. Under many circumstances, we need to know the status of a network under specific attacks. Therefore, this paper suggests defining a few commonly used network invulnerability subindexes --- $I_{0.2}$, $I_{0.5}$, $I_{0.7}$ and $I_{1.0}$, which correspond to the removal of 20\%, 50\%, 70\% and 100\% of the edges, respectively, under given attack strategies.

\section{Simulation experiments and results}
%选取的仿真网络
%     代价攻击、非代价攻击
%    不同网络性能下的测度面积
%In this section, we use three common used networks, Internet, political blogs network(Polblog), and Power grid network(Powergrid) to validate our measure. The three networks are attacked under selective attacks and random attacks, and the responding curves are plotted in fig* and fig* respectively.

To demonstrate the proposed method, we select three commonly used small-world networks as test beds: the neural network of the worm Caenorhabditis elegans (C. elegans), the power grid of the western United States (Powergrid) and the WS model \cite{16} network. The first two are real-world networks, and the WS model network is derived from the model of Watts and Strogatz. According to the WS model, we let 1000 nodes to form a ring, and for each node, link it to 10 closest right neighbors; and then let every link has a rewiring probability as $p=0.02$ to change its destination node to an arbitrary node. Any additional information of all networks is ignored except the topology. Some properties of the tested networks are listed in Table 1.

\begin{table}[htbp]
\centering
 \caption{\label{tab:table1}Properties of tested networks. $N$ is the number of nodes, $E$ is the number of edges, $\overline{D}$ is the average degree.}
 \begin{tabular}{cccc}
  \toprule
  Network & $N$ & $E$ & $\overline{D}$ \\
  \midrule
  C. elegans  & 453 & 2040 & 8.94 \\
  Powergrid  & 4941 & 6594 & 2.66 \\
  WS model  & 1000 & 10000 & 20 \\
  \bottomrule
 \end{tabular}
\end{table}

There are numberless methods to attack a system represented by network. Commonly, we concern the random failures of the systems and the malicious attacks. Therefore, this paper use  two classes of popular attack strategies \cite{146}. The first class is the random attack strategies or the random failures, i.e. , to remove the nodes or the edges randomly from a tested network. The second class is the selective attack strategies or the malicious attacks, i.e. , to remove the most important nodes or the edges one by one. Based on different definitions on the importance of the nodes or the edges, different selective attack strategies can be used \cite{152,a1}. In most circumstances, the importance used in the previous studies are the degree and the betweenness. Moreover, the calculation method of the importance can also affect the selective attack strategies, that is, we can only calculate the importance when before the attacks, termed as ``based on initial graphs'', or we can calculate the importance after every attack, termed as ``based on recalculation''. In general, the attack strategies widely used are random, ID (Initial Degree), IB (Initial Betweenness), RD (Recalculation Degree), RB (Recalculation Betweenness). Here we only choose the attack strategies based on the initial graphs, i.e., ID and IB. That is, before removing any nodes or edges, the degree or betweenness values are calculated and sorted in descending order, and then, the nodes or edges will be removed one by one according to their orders, the most important nodes or edges will be removed firstly. If some nodes or edges have the same degree or betweenness, then they will be chosen one by one randomly.

The simulation results of the proposed method under random (Rn), ID and IB attacks on edges and nodes are shown in Fig. 2 and Fig. 3, respectively, and the corresponding invulnerability indexes are shown in Tables 2 and 3.

\subsection{Simulation experiments under edge attacks}
 In these experiments, three widely used attack strategies are used, that is, the random, ID and IB edge attacks. Here, the degree of an edge \cite{146} is the product of the degrees of the linked nodes, and the betweenness of an edge \cite{146} is defined as the number of shortest paths between pairs of nodes that run along the edge.

The results of experiments on networks under random and selective attacks are shown in Fig. 2, and the corresponding indexes are listed in Table 2.

\begin{figure}

  \subfigure[Random attack]{
    \label{fig2:subfig:a} %% label for first subfigure
    \includegraphics[width=5.5cm,height=4.5cm]{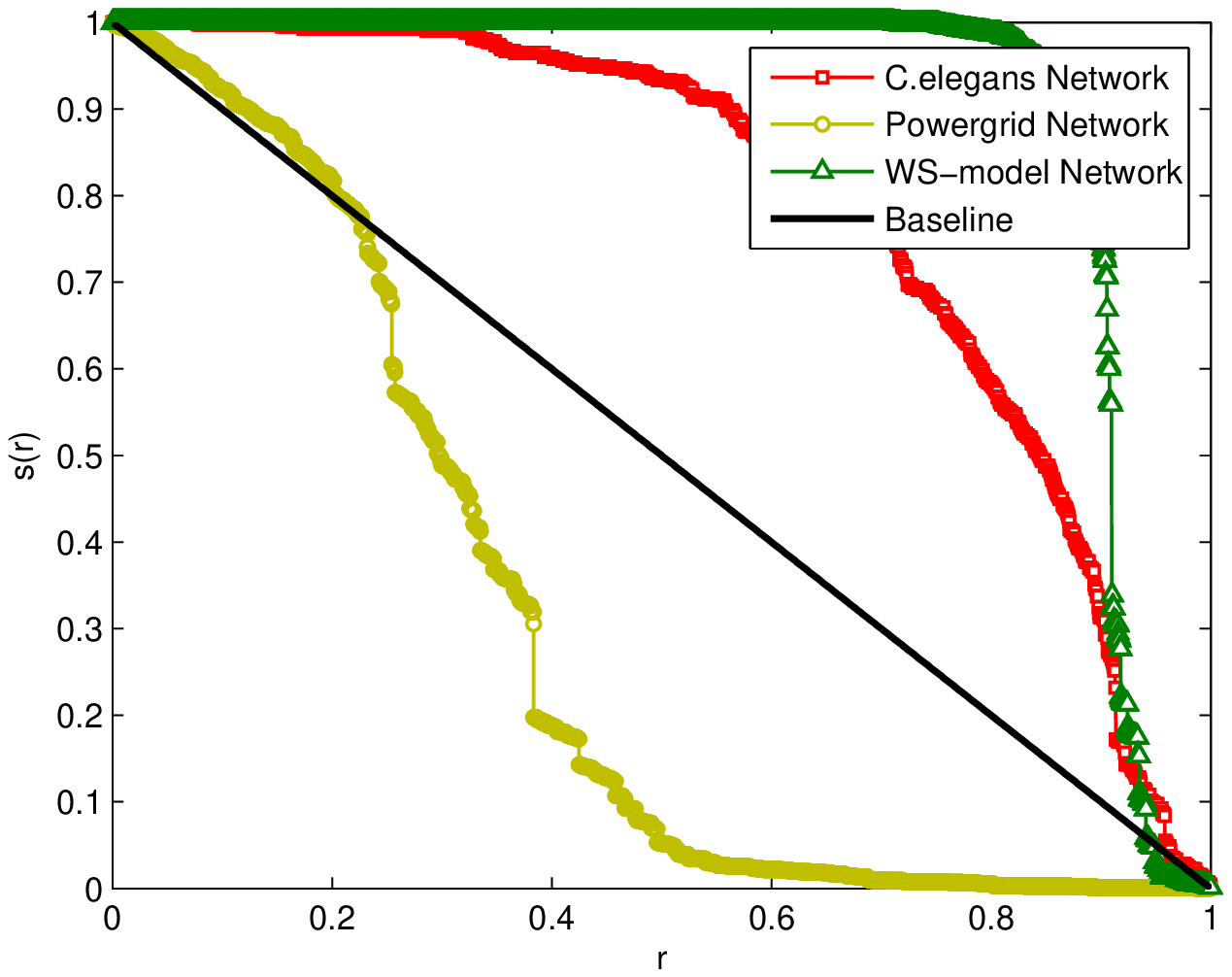}}
  \hspace{-0.8cm}
  \subfigure[ID attack]{
    \label{fig2:subfig:b} %% label for second subfigure
    \includegraphics[width=5.5cm,height=4.5cm]{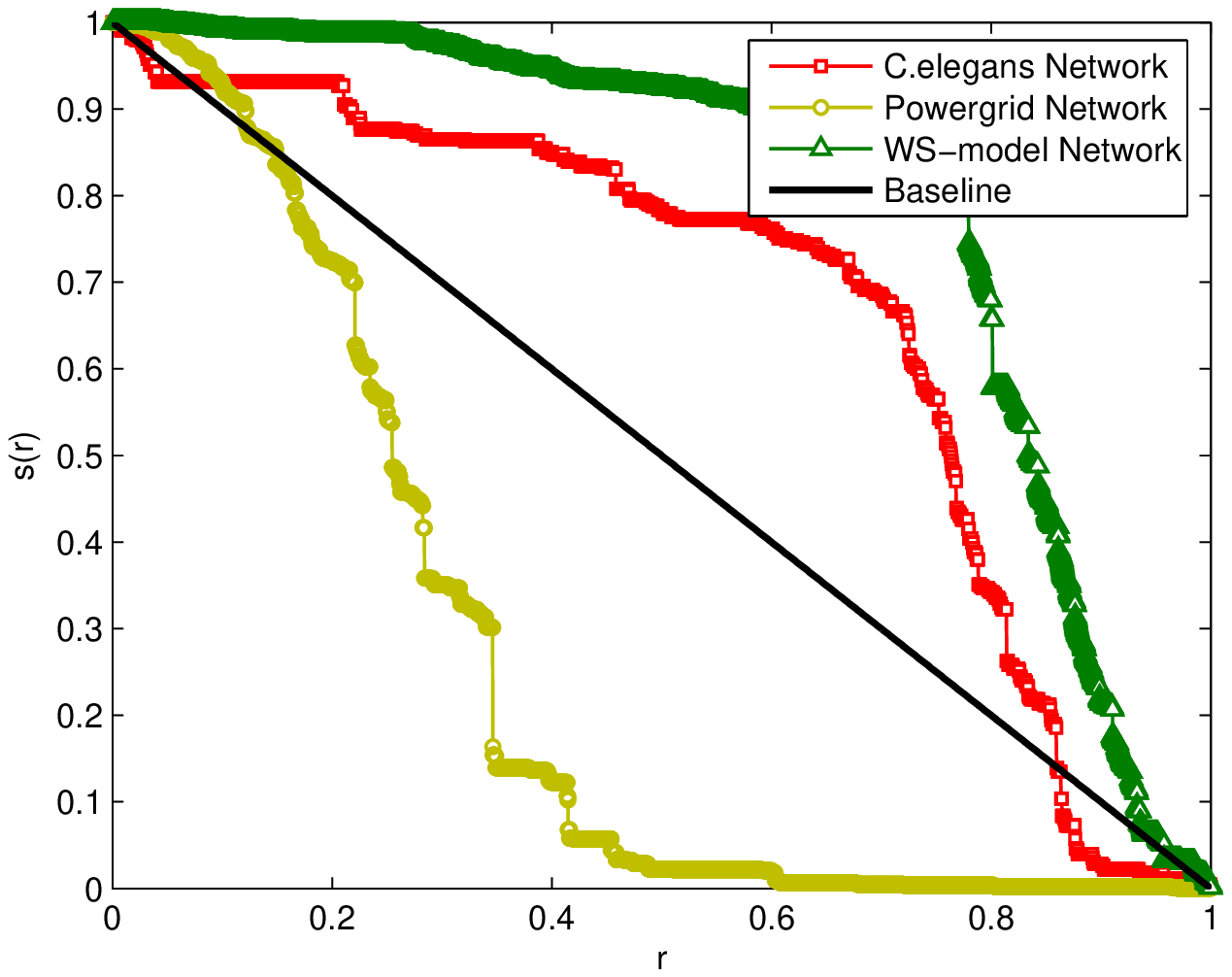}}
  \hspace{-0.8cm}
  \subfigure[IB attack]{
    \label{fig2:subfig:c} %% label for second subfigure
    \includegraphics[width=5.5cm,height=4.5cm]{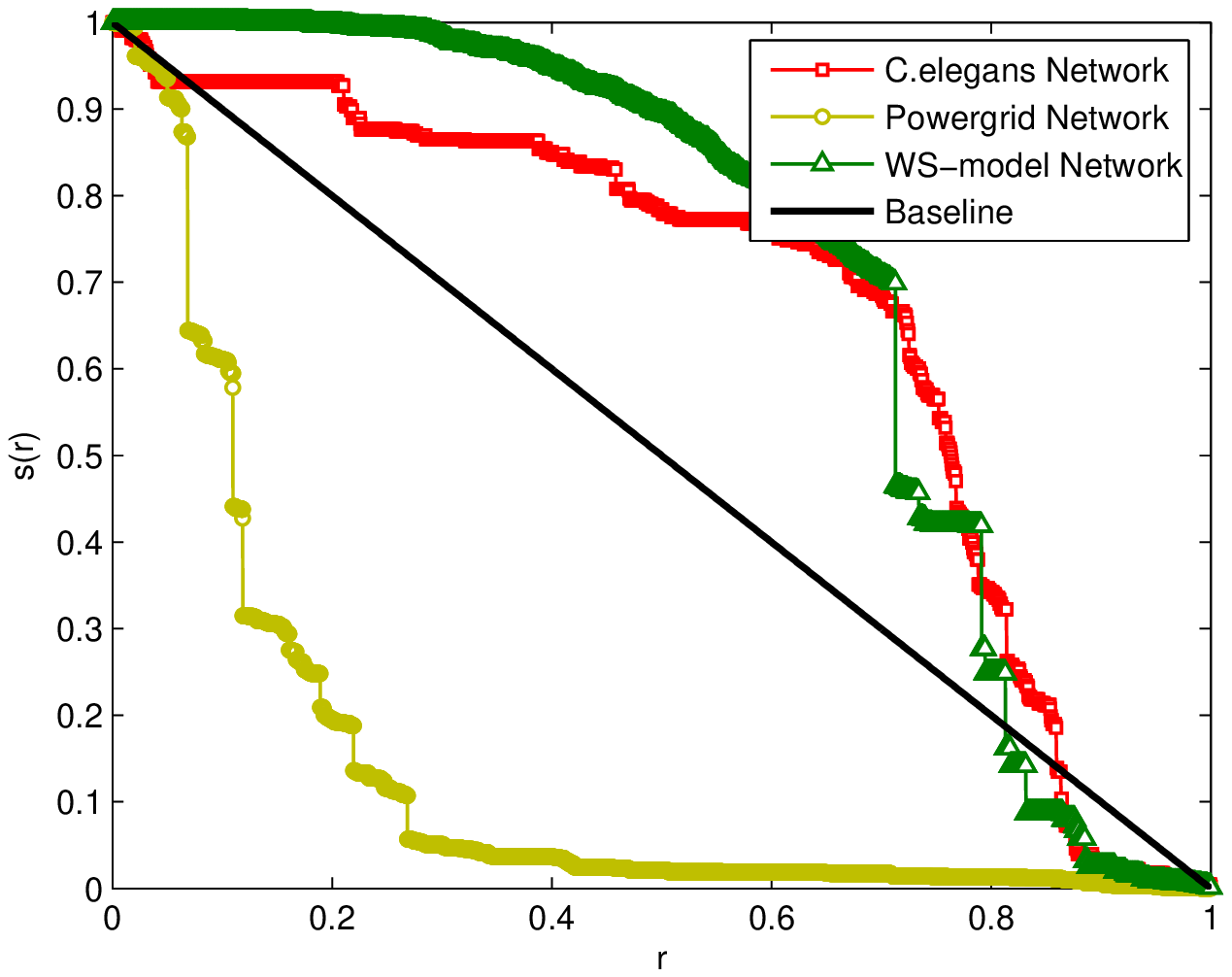}}
  \caption{Simulation experiments under edge attacks.
    $s(r)$ is the normalized network performance, which is measured by the size of the giant component, and   $r$ is the percentage of edges removed. The yellow (triangles) curves represent the Powergrid network tested under corresponding random, ID and IB attacks; the red (rectangles) curves represent the C. elegans network; the green (circles) curves represent the WS model network; and the black line is the baseline. The D in ID indicates edge degree, and B in IB indicates the edge betweenness.}
  \label{fig2:subfig} %% label for entire figure
   \end{figure}
%\begin{figure}[htbp]
%\includegraphics[width=18cm,height=6cm]{Edge_attack.eps}
%\caption{\label{fig.fig3} Attack experiments under edge attacks}

%\end{figure}

\begin{table}[htbp]
\centering
 \caption{\label{tab:table2}The invulnerability indexes of networks tested under edge attacks. }
 \begin{tabular}{cccccc}
  \toprule
  Network & Strategy & $I_{0.2}$ & $I_{0.5}$ & $I_{0.7}$ & $I_{1.0}$\\
  \midrule
  C. elegans & Rn    & 0.018 & 0.120 & 0.206 & 0.286 \\
  ---        & ID    & 0.008 & 0.070 & 0.120 & 0.166 \\
  ---        & IB    & 0.008 & 0.070 & 0.138 & 0.166 \\

   Powergrid  & Rn   & 0.008 & -0.152 & -0.302 & -0.390 \\
   ---         & ID  & 0.000 & -0.112 & -0.188 & -0.234 \\
   ---         & IB  & -0.062 & -0.240 & -0.316 & -0.358 \\

    WS model   & Rn  & 0.020 & 0.126 & 0.246 & 0.406 \\
    ---         & ID & 0.020 & 0.112 & 0.210 & 0.300 \\
    ---        & IB  & 0.020 & 0.114 & 0.194 & 0.206 \\
  \bottomrule
 \end{tabular}
\end{table}

In Fig. 2(a), most points on the network performance curves for the C. elegans (the red-rectangle curve) and the WS model network (the green-triangle curve) are clearly located above the baseline under random attacks. In contrast, a large proportion of points on the network performance curve of the Powergrid network (the yellow-circle curve) are below the baseline. Similarly, in Fig. 2(b) and (c), the relative locations of the network performance curves and the baseline do not change much, yet the network performance curves of the Powergrid (yellow-circle curve) deviate greatly from the baseline as compared with that in Fig. 2(a). Based on the definition of the baseline, we can determine that the C. elegans and WS model networks are robust and the Powergrid network is fragile under the three types of attacks.

Regarding the quantified network invulnerability (Fig. 2), the corresponding invulnerability indexes $I_{0.2}$, $I_{0.5}$ and $ I_{0.7}$ indicate changes in network invulnerability, while the invulnerability index $I_{1}$ is used to estimate network resilience. For the networks tested under the random attack strategy, the four invulnerability indexes of the WS model and C. elegans networks are positive and increase to 0.406 and 0.286, respectively. By contrast, the Powergrid becomes fragile when over 50\% edges are removed because the network's index becomes negative at $I_{0.5}$, and $I_{1}$, which is the smallest index value, equals $-0.234$. From Fig. 2 and Table 2, we can see that the WS model network is the most robust among those tested, and the Powergrid is the most fragile.

\subsection{Simulation experiments under node attacks}

The simulation results and the four corresponding invulnerability indexes are shown in Fig. 3 and Table 3, respectively.

Similar to the edge attack strategy, the random attack strategy on nodes involves removing nodes arbitrarily from the tested networks. In addition, selective attack strategies aim to remove nodes in the descending order of node degree or node betweenness based on the initial graphs. The degree of a node is the number of connections that it has to other nodes, and the betweenness of a node is defined as the number of shortest paths from all vertices to all others that pass through that node.

\begin{figure}[htbp]

  \subfigure[Random attack]{
    \label{fig3:subfig:a} %% label for first subfigure
    \includegraphics[width=5.5cm,height=4.5cm]{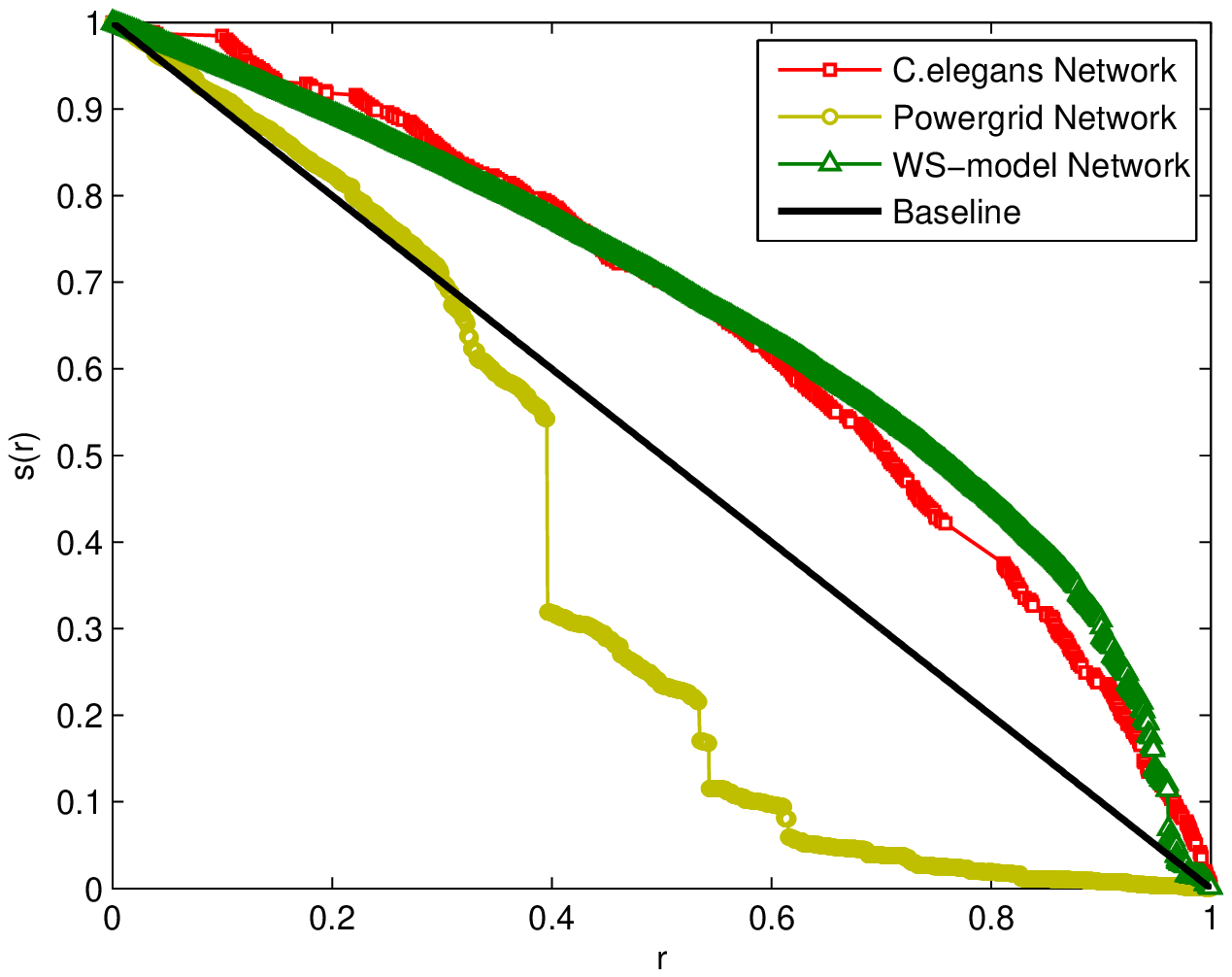}}
  \hspace{-0.8cm}
  \subfigure[ID attack]{
    \label{fig3:subfig:b} %% label for second subfigure
    \includegraphics[width=5.5cm,height=4.5cm]{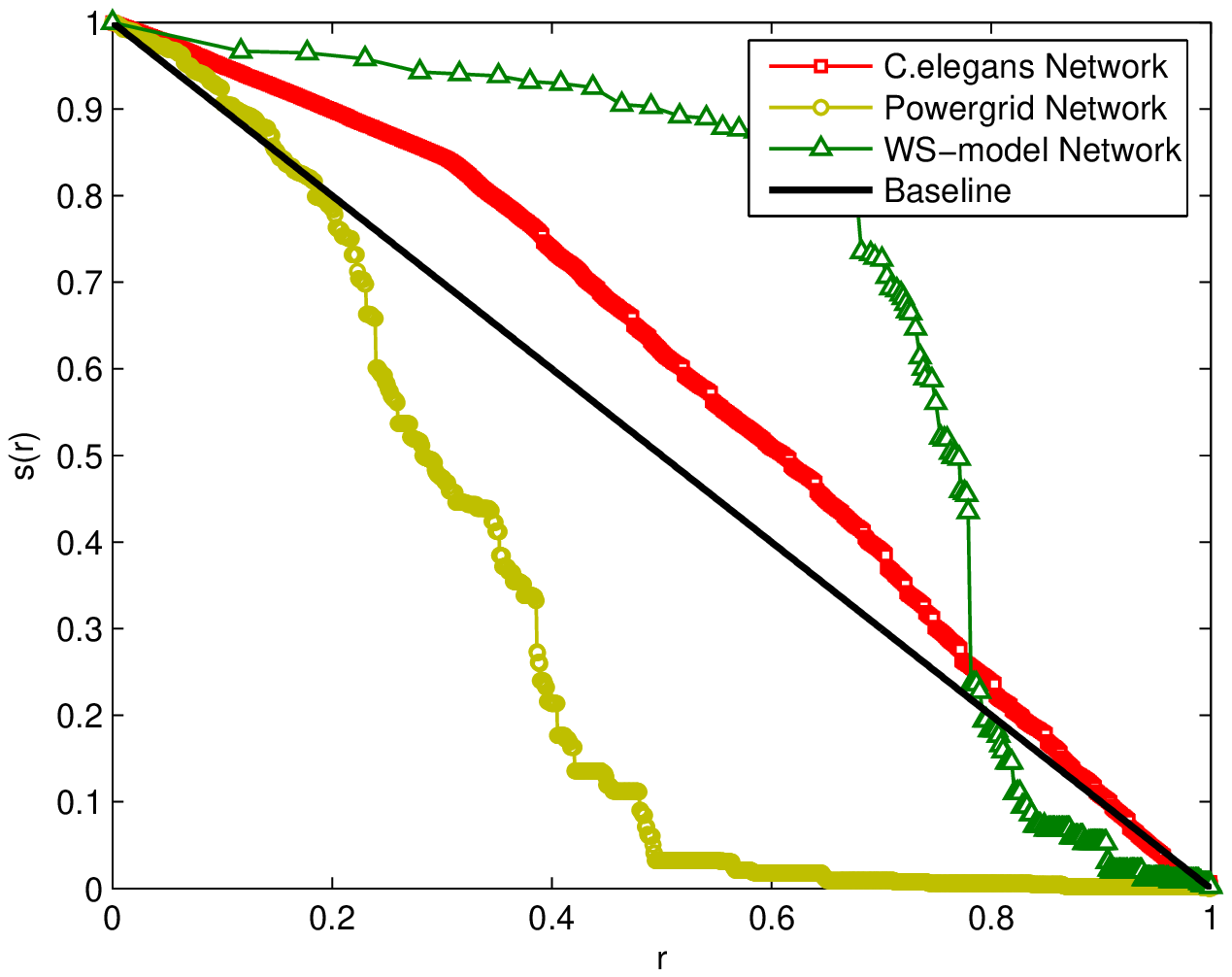}}
    \hspace{-0.8cm}
  \subfigure[IB attack]{
    \label{fig3:subfig:c} %% label for second subfigure
    \includegraphics[width=5.5cm,height=4.5cm]{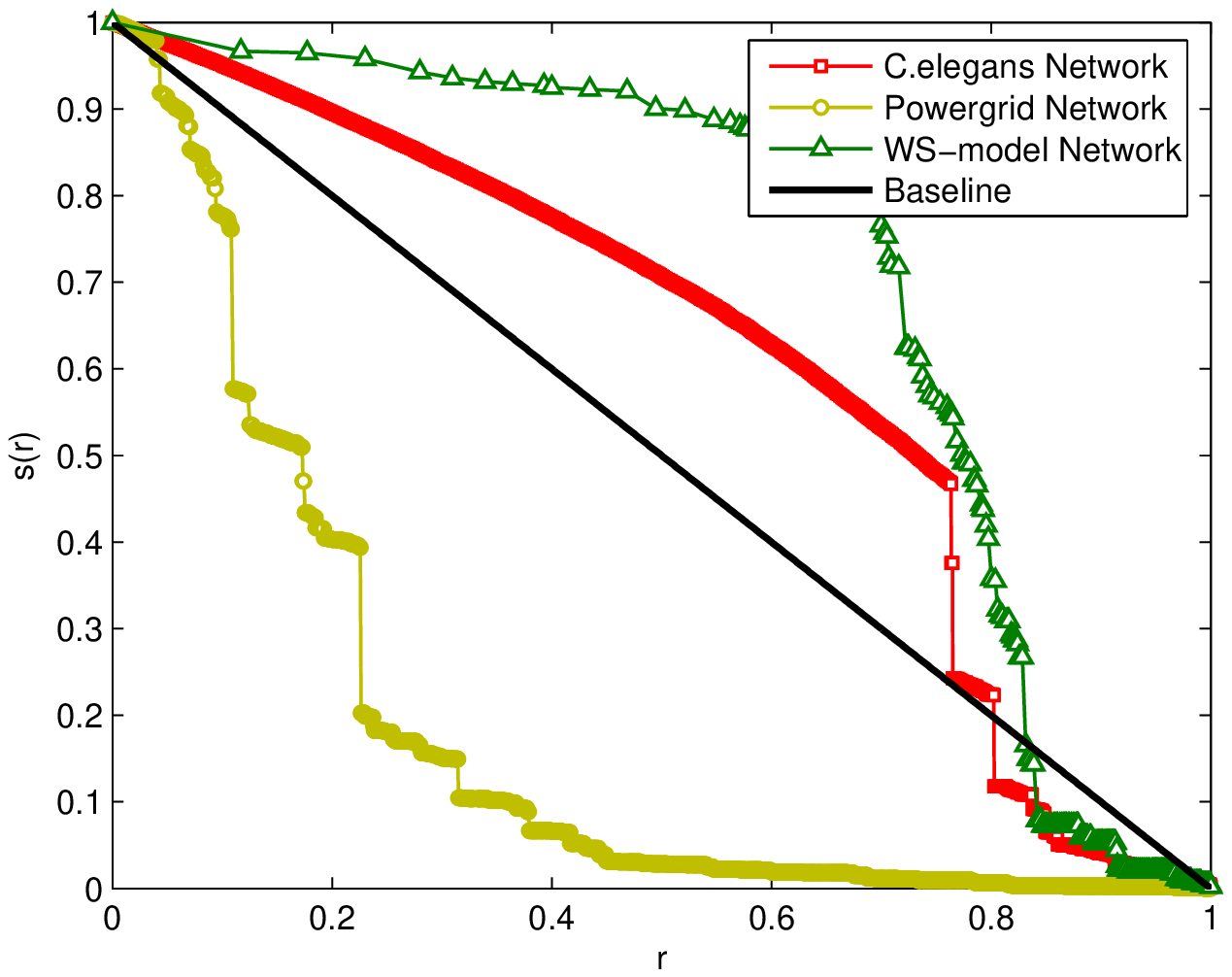}}
  \caption{Simulation experiments under node attacks. The degree and betweenness are actually the node degree and the node betweenness, respectively.}
  \label{fig3:subfig} %% label for entire figure

\end{figure}

\begin{table}[htbp]
\centering
 \caption{\label{tab:table3}The invulnerability indexes of networks tested under node attacks. }
 \begin{tabular}{cccccc}
  \toprule
  Network & Strategy & $I_{0.2}$ & $I_{0.5}$ & $I_{0.7}$ & $I_{1.0}$\\
  \midrule
  C. elegans & Rn     & 0.146 & 0.178   & 0.180 & 0.194 \\
     ---         & ID & 0.095 & 0.111   & 0.128 & 0.135 \\
    ---        & IB   & 0.103 & 0.137   & 0.154 & 0.188 \\

   Powergrid  & Rn    & -0.005 & -0.083 & -0.115 & -0.128 \\
   ---         & ID   & 0.001  & -0.094 & -0.134 & -0.156 \\
   ---         & IB   & 0.116  & -0.076 & -0.156 & -0.201\\

    WS model   & Rn   & 0.143  & 0.159 & 0.193   & 0.207 \\
   ---        & ID    & 0.119  & 0.141 & 0.189   & 0.267 \\
  ---        & IB     & 0.108  & 0.130 & 0.170   & 0.265 \\
  \bottomrule
 \end{tabular}
\end{table}

The relative locations between the network performance curves and the baseline in Fig. 3 are similar to those shown in Fig. 2, i.e., most of the points on the network performance curves of the WS model and C. elegans networks are above the baseline, and almost all of the points on the network performance curve of the Powergrid are below the baseline.

To be specific, under random attacks on the nodes (Fig. 3(a)), the network invulnerability index of the C. elegans and WS model networks satisfy the condition $I_{1}>0$, and the invulnerability of the latter is slightly larger than that of the former. The value of $I_{1}$ for the Powergrid is negative and is clearly the smallest $I_{1}$ value. From Fig. 3 and Table 3, we can see that the WS model network is the most robust among the tested networks under random attacks, and the Powergrid is the most fragile.

In addition, Fig. 3(b) and (c) have showed that the WS model network is the most robust under ID and IB attacks, meanwhile, the Powergrid is the most fragile. From Table 3, we can see that the invulnerability index of the Powergrid becomes negative when  $\alpha=0.5$. This result indicates that the network performance of the Powergrid decreases more rapidly when over 50\% of the network's edges are removed.

To validate the approximation of the index values between the node attacks and corresponding edge attacks,  we transfer the ID and IB node attacks to the ID and IB edge attacks, and compared the index values as Table 4. When transferring the node attacks to the edge attacks, for each removing node, every edge would be randomly chosen one by one to be removed. Because the sequence of the removing edges are different, the invulnerability index values of edge attacks may vary. Therefore, there are errors between the node attacks and the corresponding edge attacks. In Table 4, we carry out the node attacks and corresponding edge attacks one time for each network, and calculate the absolute errors of the invulnerability index values between under the node attacks and under the corresponding edge attacks (denoted as Error).

\begin{table}[htbp]\caption{The approximation analysis on the node attacks and corresponding edge attacks.}
\centering
 \begin{tabular}{cccccccc}
  \toprule
Network&
SubIndex&
ID Node&
ID Edge&
Error&
IB Node&
IB Edge&
Error\\
  \midrule
C. elegans & $I_{0.2}$ &0.095  &0.016  &-0.079   &0.103   &0.016   &-0.087\\
---        & $I_{0.5}$ &0.111  &0.102  &-0.092   &0.137   &0.102   &-0.036\\
---        & $I_{0.7}$ &0.128  &0.193  &0.065    &0.154   &0.193   &0.039\\
---        & $I_{1}$   &0.135  &0.211  &0.076    &0.188   &0.223   &0.035\\
Powergrid  &$I_{0.2}$  &0.000  &0.019  &0.002    &-0.062  &-0.036  &0.026\\
---        &$I_{0.5}$  &-0.112 &-0.081 &0.031    &-0.240  &-0.193  &0.047\\
---        &$I_{0.7}$  &-0.188 &-0.157 &0.031    &-0.316  &-0.269  &0.047\\
---        &$I_{1}$    &-0.234 &-0.201 &0.033    &-0.358  &-0.313  &0.045\\
WS         &$I_{0.2}$  &0.020  &0.012  &-0.077   &0.020   &0.012   &-0.077\\
---        &$I_{0.5}$  &0.112  &0.071  &-0.041   &0.114   &0.071   &-0.043\\
---        &$I_{0.7}$  &0.210  &0.128  &-0.082   &0.194   &0.129   &-0.065\\
---        &$I_{1}$    &0.300  &0.210  &-0.090   &0.206   &0.212   &0.055\\
  \bottomrule
\end{tabular}
\label{tab4}
\end{table}

From Table 4, we can see that the values are very close with the maximal absolute error less than 0.09. The results demonstrated the proposed method can assure the approximation.

From the simulation experiments discussed above, we can draw a few conclusions: 1) The proposed method can simultaneously estimate network invulnerability under both edge attacks and node attacks; 2) small-world networks can be robust under selective attacks; 3) the robustness or fragility of a specific network is consistent under node attacks and edge attacks.

\section{Conclusion}

In this paper, we proposed an invulnerability index and demonstrated a calculation method to quantify network robustness. Using small-world networks as test beds, the effectiveness of the proposed index was evaluated. The experimental results indicated that the proposed index could work under both node and edge attacks. Moreover, the invulnerability index under node attacks approximates that under edge attacks. In other words, the conclusions on the robustness or the fragility of the tested networks were consistent. Moreover, the method provided an accurate demarcation by which to distinguish the robustness from the fragility. When addressing the behavior of network performance under a series of attacks, the method also provided the critical point at which a network becomes fragile. The experimental results indicate that two of the tested networks are robust to selective and random attacks, whereas the third is fragile: i.e., some small-world networks could be robust to selective attacks.

In the future, we will use the method reported here to investigate the robustness of other types of networks under different attack strategies such as PageRank \cite{22,23} attacks and HITS \cite{24} attacks.

\ack
We thank Dr. Baobin Wang, Wenhua Du for valuable discussions, and thank Omar F. and Cecelia S. for language editing. The authors are grateful of the editors and the anonymous reviewers for improving the manuscript quality. JQ is grateful for support from the Fundamental Research Funds for the Central Universities (No. CZY12032) and Nature Science Foundation in Hubei (No. BZY11010) and the State Key Laboratory of Networking and Switching Technology (No. SKLNST-2010-1-04). BZ is grateful for support from the State Key Laboratory of Software Engineering (No. 2012-09-15) and the National Natural Science Foundation of China (No. 61273213, 61272111 and 60803095).

\section*{References}

%\bibliographystyle{unsrt}
%\bibliography{ComplexNetwork}

\end{document}